\documentclass[english,reprint, longbibliography, superscriptaddress, breaklinks=true, showkeys, showpacs=false, nofootinbib]{revtex4}
\usepackage[T1]{fontenc}
\usepackage[latin9]{inputenc}
\usepackage{color}
\usepackage{babel}
\usepackage{amsmath}
\usepackage{amssymb}
\usepackage{graphicx}
\usepackage{physics}
\usepackage[unicode=true,pdfusetitle,
 bookmarks=true,bookmarksnumbered=false,bookmarksopen=false,
 breaklinks=true,pdfborder={0 0 0},backref=false,colorlinks=true]
 {hyperref} 
\usepackage{breakurl}

\makeatletter
\@ifundefined{textcolor}{}
{%
 \definecolor{BLACK}{gray}{0}
 \definecolor{WHITE}{gray}{1}
 \definecolor{RED}{rgb}{1,0,0}
 \definecolor{GREEN}{rgb}{0,1,0}
 \definecolor{BLUE}{rgb}{0,0,1}
 \definecolor{CYAN}{cmyk}{1,0,0,0}
 \definecolor{MAGENTA}{cmyk}{0,1,0,0}
 \definecolor{YELLOW}{cmyk}{0,0,1,0}
}

\pdfoutput=1
\hypersetup{colorlinks=true,citecolor=blue,linkcolor=cyan,urlcolor=blue,filecolor= green, breaklinks=true}
\usepackage{url}
\usepackage{breakurl}

\makeatother

\begin{document}

\title{Quantitative wave-particle duality relations from the density matrix properties}

\author{Marcos L. W. Basso}
\address{Departamento de F\'isica, Centro de Ci\^encias Naturais e Exatas, Universidade Federal de Santa Maria, Avenida Roraima 1000, Santa Maria, RS, 97105-900, Brazil}

\author{Diego S. S. Chrysosthemos}

\author{Jonas Maziero}
\email{jonas.maziero@ufsm.br}

\selectlanguage{english}%

\address{Departamento de F\'isica, Centro de Ci\^encias Naturais e Exatas, Universidade Federal de Santa Maria, Avenida Roraima 1000, Santa Maria, RS, 97105-900, Brazil}

\begin{abstract}
We derive upper bounds for Hilbert-Schmidt's quantum coherence of general states of a $d$-level quantum system, a qudit,  in terms of its incoherent uncertainty, with the latter quantified using the linear and von Neumann's entropies of the corresponding closest incoherent state. Similar bounds are obtained for Wigner-Yanase's coherence. The reported inequalities are also given as coherence-populations trade-off relations. 
As an application example of these inequalities, we derive quantitative wave-particle duality relations for multi-slit interferometry. Our framework leads to the identification of predictability measures complementary to Hilbert-Schmidt's, Wigner-Yanase's, and $l_{1}$-norm quantum coherences. The quantifiers reported here for the wave and particle aspects of a quanton follow directly from the defining properties of the quantum density matrix (i.e., semi-positivity and unit trace), contrasting thus with most related results from the literature.
\end{abstract}


\maketitle

\section{Introduction}

\label{intro}

Quantum Information Science (QIS) is a rapidly developing interdisciplinary
field harnessing and instigating some of the most advanced results
in physics, information theory, computer science, mathematics, material
science, engineering, and artificial intelligence \cite{fallani,simmons,pirandola,schleier,iop,fitzsimons,adesso,
stuhler,preskill,cowen,biamonte}.
Nowadays we know of several aspects of quantum systems that contrast
them from the classical ones. Of particular interest in QIS is to
investigate how these quantum features can be harnessed to devise
more efficient protocols for information processing, transmission,
storage, acquisition, and protection.

A few examples of connections among quantum features and advantages
in QIS protocols are as follows. Of special relevance for the whole
of QIS, and in particular for the development of a quantum internet,
is the use of quantum entanglement as a channel for quantum teleportation
\cite{popescu,cavalcanti}. As for one of the most advanced branches
of QIS, its is known that quantum nonlocality and quantum steering
are needed for device independent and semi-independent quantum communication,
respectively \cite{brunner,branciard}. By its turn, the use of quantum
squeezing and quantum discord was related with increased precision
of measurements in quantum metrology \cite{schnabel,girolami}. Besides,
quantum contextuality and quantum coherence (QC) were connected with
the speedup of some algorithms in quantum computation \cite{howard,shi}.

Quantum coherence, a kind of quantum superposition \cite{theurer},
is directly related to the existence of incompatible observables in
Quantum Mechanics; and is somewhat connected to most of the quantumnesses
mentioned above. Therefore, it is a natural research program trying
to understand QC from several perspectives. Recently, researchers
have been developing a resource theory framework to quantify QC, the
so called resource theory of coherence (see e.g. Refs. \cite{aberg,prillwitz,streltsov,chitambar_jpa}
and references therein). In this resource theory, given an orthonormal
reference basis $\{|\beta_{n}\rangle\}_{n=1}^{d}$, with $d=\dim\mathcal{H}$, for
a system with state space $\mathcal{H}$, the free states are incoherent
mixtures of these base states:
\begin{equation}
\iota=\sum_{n=1}^{d}\iota_{n}|\beta_{n}\rangle\langle\beta_{n}|,\label{eq:iota}
\end{equation}
where $\{|\iota_{n}\rangle\}_{n=1}^{d}$ is a probability
distribution. A geometrical way of defining functions to quantify
coherence is via the minimum distance from $\rho$ to incoherent states:
\begin{equation}
C_{D}(\rho)=\min_{\iota}D(\rho,\iota)=:D(\rho,\iota_{\rho}^{D}),
\end{equation}
where $\iota_{\rho}^{D}$ is the closest incoherent state to $\rho$
under the distance measure $D$. If $C_{D}$ does not increase under
incoherent operations, which are those quantum operations mapping
incoherent states to incoherent states, then its dubbed a coherence
monotone.

In Quantum Mechanics \cite{messiah}, the more general description
of a system state is given by its density operator $\rho=\sum_{m}p_{m}|\psi_{m}\rangle\langle\psi_{m}|$,
where $\{p_{m}\}$ is a probability distribution and $\{|\psi_{m}\rangle\}$
are state vectors \cite{fano}. Because of this ensemble interpretation,
the density operator is required to be a positive (semi-definite)
linear operator, besides having trace equal to one \cite{wilde}.
If we consider, for instance, two-level systems whose density operator
represented in the orthonormal basis $\{|\beta_{n}\rangle\}_{n=1}^{2}$
reads
\begin{equation}
\rho=\begin{bmatrix}\langle\beta_{1}|\rho|\beta_{1}\rangle & \langle\beta_{1}|\rho|\beta_{2}\rangle\\
\langle\beta_{2}|\rho|\beta_{1}\rangle & \langle\beta_{2}|\rho|\beta_{2}\rangle
\end{bmatrix}=:\begin{bmatrix}\rho_{1,1} & \rho_{1,2}\\
\rho_{1,2}^{*} & 1-\rho_{1,1}
\end{bmatrix},
\end{equation}
these properties impose a well known restriction on the off-diagonal
elements of $\rho$, its coherences, by the product of its diagonal
elements, its populations:
\begin{equation}
\rho_{1,1}(1-\rho_{1,1})\ge|\rho_{1,2}|^{2}.\label{eq:tos}
\end{equation}
The product of the populations of $\rho$ can be seen as the \emph{incoherent
uncertainty} we have about measurements of an observable with eigenvectors
$\{|\beta_{n}\rangle\}_{n=1}^{2}$, since it is independent of the
whole ensemble coherences. On the other hand, the presence of non-null
off-diagonal elements of $\rho$ implies that one or more members
of the ensemble are a coherent superposition of the base states $\{|\beta_{n}\rangle\}_{n=1}^{2}$.

It is an interesting mathematical, physical, and possibly practical
problem to derive quantum coherence--incoherent uncertainty trade-off
relations regarding general-discrete quantum systems. In this article,
we obtain such trade-offs for one-qudit ($d$-level) quantum systems.
We start considering Hilbert-Schmidt's coherence (HSC)
function \cite{maziero}, that has a convenient algebraic
structure but is known not to be a coherence monotone \cite{baumgratz}.
We also regard Wigner-Yanase's coherence (WYC) \cite{yu_Cwy},
which is a coherence monotone. To quantify the incoherent uncertainty of a state $\rho$, we employ linear entropy and von Neumann's entropy of its closest
incoherent state or of the diagonal of $\sqrt{\rho}$. It is worthwhile mentioning that the relation between
``quantum'' and ``classical'' uncertainties, and other complementarity
relations, have been investigated in other contexts elsewhere \cite{puchala,korzekwa,bagan,singh,cheng,liu}.

Wave-particle duality has been at the center stage of conceptual discussions in Quantum Mechanics since
its early days \cite{pessoajr}. Recently, several authors have investigated about possible definitions of predictability and visibility quantifiers for $d$-slits interferometers. For a review of the literature, see e.g. Ref. \cite{qureshi}. D\"urr's \cite{dur} and Englert et al.'s \cite{englert} criteria can be taken as a standard for checking for the reliability of newly defined predictability measures $P(\rho)$ and interference pattern visibility quantifiers $W(\rho)$. For $P$, these required properties can be restated as follows:
\begin{itemize}
\item[P1] $P$ must be a continuous function of the diagonal elements of the density matrix.
\item[P2] $P$ must be invariant under permutations of the paths' indexes.
\item[P3] If $\rho_{j,j}=1$ for some $j$, then $P$ must reach its maximum value.
\item[P4] If $\{\rho_{j,j}=1/d\}_{j=1}^{d}$, then $P$ must reach its minimum value.
\item[P5] If $\rho_{j,j}>\rho_{k,k}$ for some $(j,k)$, the value of $P$ cannot be increased by setting $\rho_{j,j}\rightarrow\rho_{j,j}-\epsilon$ and $\rho_{k,k}\rightarrow\rho_{k,k}+\epsilon$, for $\epsilon\in\mathbb{R}_{+}$ and $\epsilon\ll1$.
\item[P6] $P$ must be a convex function, i.e., $P(\omega\xi+(1-\omega)\eta)\le \omega P(\xi)+(1-\omega)P(\eta)$, for $0\le \omega\le 1$ and for $\xi$ and $\eta$ being valid density matrices.  
\end{itemize}

One can also write down a list of required properties for the functions to be used to quantify the wave aspect $W$ of a quanton in a $d$-slit interferometer \citep{dur,englert}:
\begin{itemize}
\item[W1] $W$ must be a continuous function of the elements of the density matrix.
\item[W2] $W$ must be invariant under permutations of the paths' indexes.
\item[W3] If $\rho_{j,j}=1$ for some $j$, then $W$ must reach its minimum value.
\item[W4] If $\rho$ is a pure state and $\{\rho_{j,j}=1/d\}_{j=1}^{d}$, then $W$ must reach its maximum value.
\item[W5] $W$ cannot be increased when decreasing $|\rho_{j,k}|$ by an infinitesimal amount, for $j\ne k$.
\item[W6] $W$ must be a convex function, i.e., $W(\omega\xi+(1-\omega)\eta)\le \omega W(\xi)+(1-\omega)W(\eta)$, for $0\le \omega\le 1$ and $\xi$ and $\eta$ are well defined density matrices.  
\end{itemize} 

There are convincing arguments indicating quantum coherence as a good measure for the wave aspect of a quanton \cite{qureshiC}. And we'll show that our coherence-incoherent uncertainty trade-off relations can be applied to obtain quantitative wave-particle duality relations, with the associated predictability and visibility measures satisfying the criteria listed above. The framework devised in this article shows that the semi-positiveness and unit trace of the quantum density matrix leads to the identification of predictability measures complementary to HSC, to WYC, and to  $l_{1}$-norm quantum coherence (L1C).

We organized the remainder of this article in the following manner.
In Sec. \ref{sec:gm}, we present the Gell-Mann matrix basis (GMB),
defined using any vector basis of $\mathbb{C}^{d}$, and discuss the
representation of general and diagonal matrices in the GMB. We obtain
trade-off relations between quantum coherence and incoherent uncertainty
measuring the first with Hilbert-Schmidt coherence in Sec. \ref{sec:tohs}. We obtain analogous relations using Wigner-Yanase's coherence in
Sec. \ref{sec:tohe}. We show how to write these inequalities as coherence-populations
trade-off relations in Sec. \ref{sc:hscp}. In Sec. \ref{sec:hsc} we report quantitative visibility-predictability
duality relations identified directly from our coherence-incoherent uncertainty inequalities.  In Secs. \ref{sec:wyc} and \ref{sec:l1c} we apply the positivity and unit trace of the density matrix to identify predictability measures complementary to WYC and L1C, respectively. We give our conclusions in Sec. \ref{sec:conc}.

\section{Gell-Mann basis for $\mathbb{C}^{d\mathrm{x}d}$}

\label{sec:gm}

Let $\{|\beta_{m}\rangle\}_{m=1}^{d}$ be any given vector basis for
$\mathbb{C}^{d}$. Using this basis, we can define the generalized
Gell-Mann's matrices as \cite{bertlmann}:
\begin{eqnarray}
\Gamma_{j}^{d} & := & \sqrt{\frac{2}{j(j+1)}}\sum_{m=1}^{j+1}(-j)^{\delta_{m,j+1}}|\beta_{m}\rangle\langle\beta_{m}|,\\
\Gamma_{k,l}^{s} & := & |\beta_{k}\rangle\langle\beta_{l}|+|\beta_{l}\rangle\langle\beta_{k}|,\\
\Gamma_{k,l}^{a} & := & -i(|\beta_{k}\rangle\langle\beta_{l}|-|\beta_{l}\rangle\langle\beta_{k}|),
\end{eqnarray}
where, if not stated otherwise, we use the following possible values for the indexes $j,k,l$ up to Sec. \ref{sc:hscp} of this article:
\begin{equation}
j=1,\cdots,d-1\mbox{ and }1\le k<l\le d.
\end{equation}
One can easily see that these matrices are Hermitian and traceless.
Besides, if we use $\Gamma_{0}^{d}$ for the $d \times d$ identity
matrix, it is not difficult to verify that under the Hilbert-Schmidt's
inner product,
\begin{equation}
\langle A|B\rangle_{hs}:=\mathrm{Tr}(A^{\dagger}B),
\end{equation}
with $A,B\in\mathbb{C}^{d \times d}$, the set
\begin{equation}
\left\{ \frac{\Gamma_{0}^{d}}{\sqrt{d}},\frac{\Gamma_{j}^{d}}{\sqrt{2}},\frac{\Gamma_{k,l}^{\tau}}{\sqrt{2}}\right\} ,
\end{equation}
with $\tau=s,a$, forms an orthonormal basis for $\mathbb{C}^{d \times d}$
\cite{bertlmann}. So, any matrix $X\in\mathbb{C}^{d \times d}$
can be decomposed in this basis, called hereafter of Gell-Mann's basis
(GMB), as follows:
\begin{equation}
X=\frac{\mathrm{Tr}(X)}{d}\Gamma_{0}^{d}+\frac{1}{2}\sum_{j}\langle\Gamma_{j}^{d}|X\rangle\Gamma_{j}^{d}+\frac{1}{2}\sum_{k,l,\tau}\langle\Gamma_{k,l}^{\tau}|X\rangle\Gamma_{k,l}^{\tau}.
\end{equation}
We observe that the most general decomposition in GMB of a matrix
$X_{d}\in\mathbb{C}^{d \times d}$ which is diagonal in the basis
$\{|\beta_{m}\rangle\}_{m=1}^{d}$ shall be given by:
\begin{equation}
X_{d}=\frac{\mathrm{Tr}(X_{d})}{d}\Gamma_{0}^{d}+\frac{1}{2}\sum_{j}\langle\Gamma_{j}^{d}|X_{d}\rangle\Gamma_{j}^{d},
\end{equation}
i.e., only the diagonal elements of the GMB can have non-null components
in this decomposition.

\section{Trade-off relations between quantum coherence and incoherent uncertainty}

\subsection{Upper bound for Hilbert-Schmidt's coherence}

\label{sec:tohs}

The Hilbert-Schmidt's coherence (HSC) of a quantum state $\rho$ is
defined as \cite{maziero}
\begin{equation}
C_{hs}(\rho):=\min_{\iota}||\rho-\iota||_{hs}^{2},\label{eq:hsc}
\end{equation}
with the Hilbert-Schmidt's norm of a matrix $A\in\mathbb{C}^{d \times d}$
being defined as
\begin{equation}
||A||_{hs}:=\sqrt{\langle A|A\rangle_{hs}},
\end{equation}
and here the minimization is taken over the incoherent states of Eq.
(\ref{eq:iota}). For general one-qudit states, using the decompositions
in GMB:
\begin{eqnarray}
\rho & = & \frac{1}{d}\Gamma_{0}^{d}+\frac{1}{2}\sum_{j}\langle\Gamma_{j}^{d}|\rho\rangle\Gamma_{j}^{d}+\frac{1}{2}\sum_{k,l,\tau}\langle\Gamma_{k,l}^{\tau}|\rho\rangle\Gamma_{k,l}^{\tau},\label{eq:rho}\\
\iota & = & \frac{1}{d}\Gamma_{0}^{d}+\frac{1}{2}\sum_{j}\langle\Gamma_{j}^{d}|\iota\rangle\Gamma_{j}^{d},
\end{eqnarray}
the analytical formulas for the HSC and for the associated closest
incoherent state were obtained in Ref. \cite{maziero} and read, respectively:
\begin{eqnarray}
C_{hs}(\rho) & = & \frac{1}{2}\sum_{k,l,\tau}\langle\Gamma_{k,l}^{\tau}|\rho\rangle^{2},\label{eq:hsc-1}\\
\iota_{\rho}^{hs} & = & \frac{1}{d}\Gamma_{0}^{d}+\frac{1}{2}\sum_{j}\langle\Gamma_{j}^{d}|\rho\rangle\Gamma_{j}^{d}.\label{eq:iotahs}
\end{eqnarray}

The main tool we use to obtain some of the results reported in this article
is a condition for matrix positivity. The eigenvalues of a matrix
$A\in\mathbb{C}^{d \times d}$, let us call them $a$, can be obtained
from \cite{kuttler}: 
\begin{eqnarray}
0 & = & \det(A-a\Gamma_{0}^{d})\\
 & = & \sum_{(j_{1},j_{2},\cdots,j_{d})}\mathrm{sgn}_{d}(j_{1},j_{2},\cdots,j_{d})(A_{1,j_{1}}-a\delta_{1,j_{1}})(A_{2,j_{2}}-a\delta_{2,j_{2}})\cdots(A_{d,j_{d}}-a\delta_{d,j_{d}})\\
 & = & (-1)^{d}c_{d}a^{d}+(-1)^{d-1}c_{d-1}a^{d-1}+(-1)^{d-2}c_{d-2}a^{d-2}+\cdots+c_{2}a^{2}-c_{1}a+c_{0}.\label{eq:coeff}
\end{eqnarray}
By Descartes rule of signs (see e.g. Ref. \cite{avendano} and references
therein), we see that for $A$ to be a positive matrix, we have to
have non-negativity for all the coefficients $\{c_{m}\ge0\}_{m=0}^{d}$.
In this article, we shall look at the positivity of:
\begin{eqnarray}
c_{d-2} & = & \sum_{(j_{1},j_{2})}\mathrm{sgn}_{d}(j_{1},j_{2},3\cdots,d)A_{1,j_{1}}A_{2,j_{2}}+\cdots+\sum_{(j_{1},j_{d})}\mathrm{sgn}_{d}(j_{1},2,\cdots,d-1,j_{d})A_{1,j_{1}}A_{d,j_{d}}\nonumber \\
 &  & +\sum_{(j_{2},j_{3})}\mathrm{sgn}_{d}(1,j_{2},j_{3},4,\cdots,d)A_{2,j_{2}}A_{3,j_{3}}+\cdots+\sum_{(j_{2},j_{d})}\mathrm{sgn}_{d}(1,j_{2},3,\cdots,d-1,j_{d})A_{2,j_{2}}A_{d,j_{d}}\nonumber \\
 &  & +\cdots+\sum_{(j_{d-1},j_{d})}\mathrm{sgn}_{d}(1,2,\cdots d-2,j_{d-1},j_{d})A_{d-1,j_{d-1}}A_{d,j_{d}}\\
 & = & \sum_{m=1}^{d-1}\sum_{n=m+1}^{d}(A_{m,m}A_{n,n}-A_{m,n}A_{n,m})\\
 & = & \frac{1}{2}\left(\left(\mathrm{Tr}(A)\right)^{2}-\mathrm{Tr}\left(A^{2}\right)\right)\ge0.\label{eq:cdm2}
\end{eqnarray}

Using the orthonormality of GMB, i.e., the inner product between different
elements of the GMB is zero and $\langle\Gamma_{0}^{d}|\Gamma_{0}^{d}\rangle=d$
and $\langle\Gamma_{j}^{d}|\Gamma_{j}^{d}\rangle=\langle\Gamma_{k,l}^{\tau}|\Gamma_{k,l}^{\tau}\rangle=2$,
the positivity condition for the coefficient in Eq. (\ref{eq:cdm2})
applied to the density matrix of Eq. (\ref{eq:rho}), as $\mathrm{Tr}(\rho)=1$,
can be rewritten as
\begin{eqnarray}
0 & \le & 1-\mathrm{Tr}(\rho^{2})\\
 & = & 1-\frac{1}{d}-\frac{1}{2}\sum_{j}\langle\Gamma_{j}^{d}|\rho\rangle^{2}-\frac{1}{2}\sum_{k,l,\tau}\langle\Gamma_{k,l}^{\tau}|\rho\rangle^{2}.
\end{eqnarray}
Now, if we use the formula for the HSC in Eq. (\ref{eq:hsc-1}), this
inequality can be cast as a restriction to the HSC:
\begin{equation}
C_{hs}(\rho)\le\frac{d-1}{d}-\frac{1}{2}\sum_{j}\langle\Gamma_{j}^{d}|\rho\rangle^{2}.\label{eq:hsto1}
\end{equation}

\begin{figure}
\includegraphics[scale=0.92]{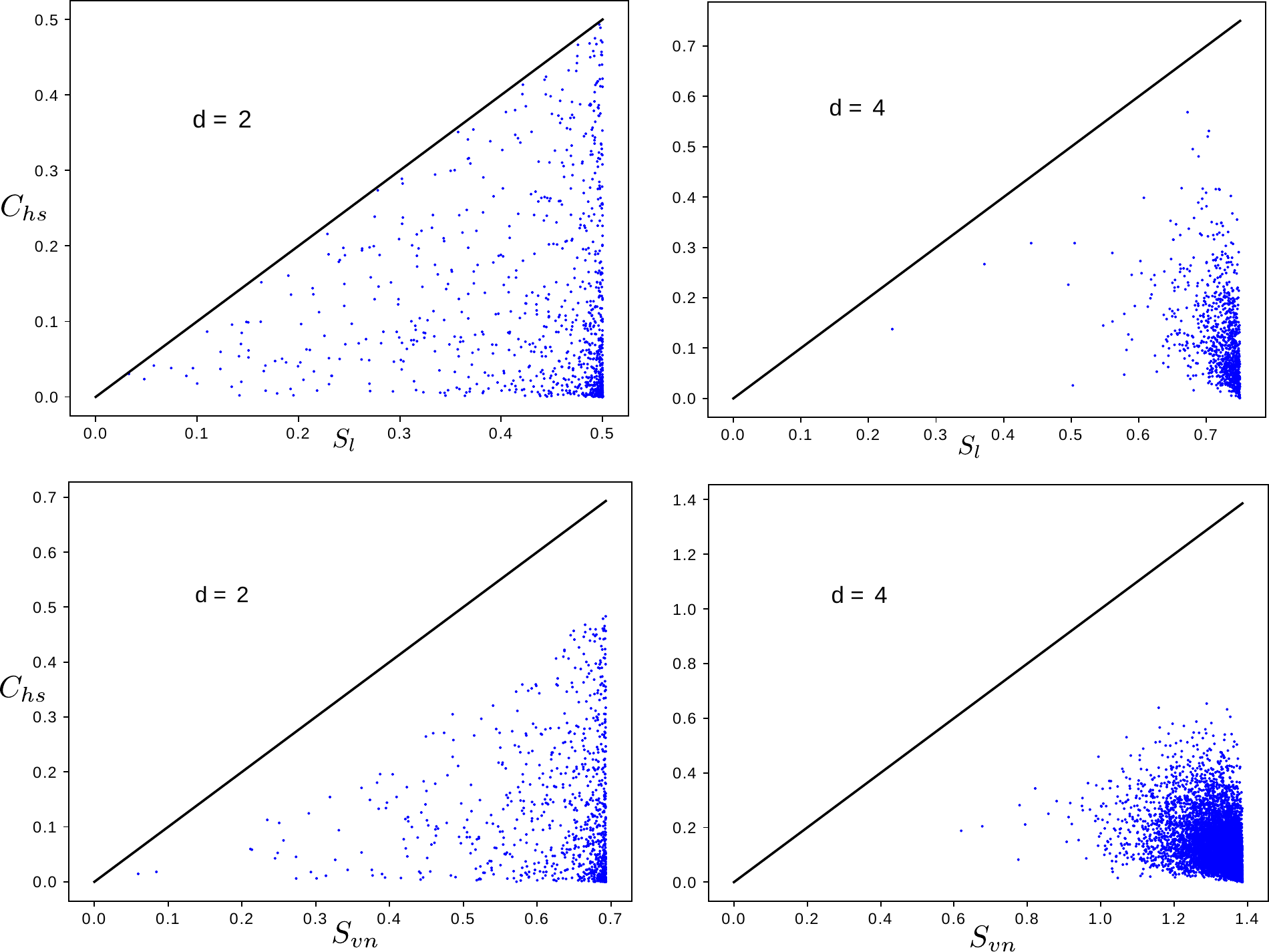}
\caption{(color online) ``Verification'' of the Hilbert-Schmidt quantum coherence--incoherent
uncertainty trade-off relations of Eq. (\ref{eq:tohs}) for one thousand
random density matrices generated for each value of the system dimension
$d$. The random states were created using the standard method described
in Refs. \cite{maziero_rrho,maziero_ft}. The $y$ axis is for $C_{hs}(\rho)$
and the $x$ axis is for $S_{l}(\iota_{\rho}^{hs})$ or $S_{vn}(\iota_{\rho}^{hs})$,
with $\iota_{\rho}^{hs}$ given in Eq. (\ref{eq:iotahs}). The black
lines stand for $C_{hs}(\rho)=S_{l}(\iota_{\rho}^{hs})$ and for $C_{hs}(\rho)=S_{vn}(\iota_{\rho}^{hs})$.}
\label{fig:hs}
\end{figure}

If we utilize again the orthonormality of the GMB, the right hand
side of this inequality is easily seen to be the incoherent uncertainty
of the state $\rho$ measured using the linear entropy of its closest
incoherent state (Eq. (\ref{eq:iotahs})), i.e.,
\begin{equation}
S_{l}(\iota_{\rho}^{hs})=1-\mathrm{Tr}\left(\left(\iota_{\rho}^{hs}\right)^{2}\right)=\frac{d-1}{d}-\frac{1}{2}\sum_{j}\langle\Gamma_{j}^{d}|\rho\rangle^{2}.
\end{equation}
Now, using $-\ln x\ge1-x$ \cite{nielsen}, we can get an upper bound
for the linear entropy in terms of von Neumann's entropy as follows:
\begin{eqnarray}
S_{vn}(x) & := & \mathrm{Tr}(x(-\ln x))\\
 & \ge & \mathrm{Tr}(x(1-x))=\mathrm{Tr}(x)-\mathrm{Tr}(x^{2})=S_{l}(x)+\mathrm{Tr}(x)-1.
\end{eqnarray}
Gathering the results above, as $\mathrm{Tr}(\iota_{\rho}^{hs})=1$,
we have obtained the following quantum coherence--incoherent uncertainty
trade-off relations:
\begin{equation}
C_{hs}(\rho)\le S_{l}(\iota_{\rho}^{hs})\le S_{vn}(\iota_{\rho}^{hs}),\label{eq:tohs}
\end{equation}
which are valid for any one-qudit state. The ``verification'' of these entropic
inequalities using random states is presented in Fig. \ref{fig:hs}.
We observe that the upper bound given by linear entropy is tight for
qubits ($d=2$). However, as the dimension increases, and typicality
is approached \cite{ledoux}, the upper bounds get less and less tight.

\subsection{Upper bound for Wigner-Yanase's coherence}
\label{sec:tohe}

In this subsection we shall deal with Wigner-Yanase's coherence (WYC) \cite{yu_Cwy}:
\begin{align}
C_{wy}(\rho)  & := \sum_{j=1}^{d}I_{wy}(\rho,|\beta_{j}\rangle\langle\beta_{j}|):=-\frac{1}{2}\sum_{j=1}^{d}\mathrm{Tr}(([\sqrt{\rho},|\beta_{j}\rangle\langle\beta_{j}|])^{2})  \\
& = \frac{1}{2}\sum_{j=1}^{d}\left(\langle\beta_{j}|\rho|\beta_{j}\rangle+\sum_{k=1}^{d}|\langle\beta_{j}|\sqrt{\rho}|\beta_{k}\rangle|^{2}-2\langle\beta_{j}|\sqrt{\rho}|\beta_{j}\rangle^{2}\right),
\end{align}
with $[\cdot,\cdot]$ being the commutator. Identifying the following relation between a diagonal element of $\rho$ and the elements in the corresponding row of $\sqrt{\rho}$:
\begin{equation}
    \langle\beta_{j}|\rho|\beta_{j}\rangle = \sum_{k=1}^{d}\langle\beta_{j}|\sqrt{\rho}|\beta_{k}\rangle\langle\beta_{k}|\sqrt{\rho}|\beta_{j}\rangle = \sum_{k=1}^{d}|\langle\beta_{j}|\sqrt{\rho}|\beta_{k}\rangle|^{2},
\end{equation}
we can write
\begin{align}
    C_{wy}(\rho) & = \sum_{j=1}^{d}\sum_{k=1}^{d}|\langle\beta_{j}|\sqrt{\rho}|\beta_{k}\rangle|^{2} - \sum_{j=1}^{d}\langle\beta_{j}|\sqrt{\rho}|\beta_{j}\rangle^{2}  \\
    & = \sum_{j\ne k}|\langle\beta_{j}|\sqrt{\rho}|\beta_{k}\rangle|^{2}.
\end{align}
Now, by using $\langle\Gamma_{k,l}^{s}|\sqrt{\rho}\rangle=2\Re((\sqrt{\rho})_{k,l})$ and $\langle\Gamma_{k,l}^{a}|\sqrt{\rho}\rangle=-2\Im((\sqrt{\rho})_{k,l})$, where $\Re((\sqrt{\rho})_{k,l})$ and $\Im((\sqrt{\rho})_{k,l})$ are the real and imaginary part of $(\sqrt{\rho})_{k,l}$ respectively, we get
\begin{equation}
     C_{wy}(\rho)=\frac{1}{2}\sum_{k,l,\tau}\langle\Gamma_{k,l}^{\tau}|\sqrt{\rho}\rangle^{2}. \label{eq:wyc}
\end{equation}

\begin{figure}
\includegraphics[scale=0.85]{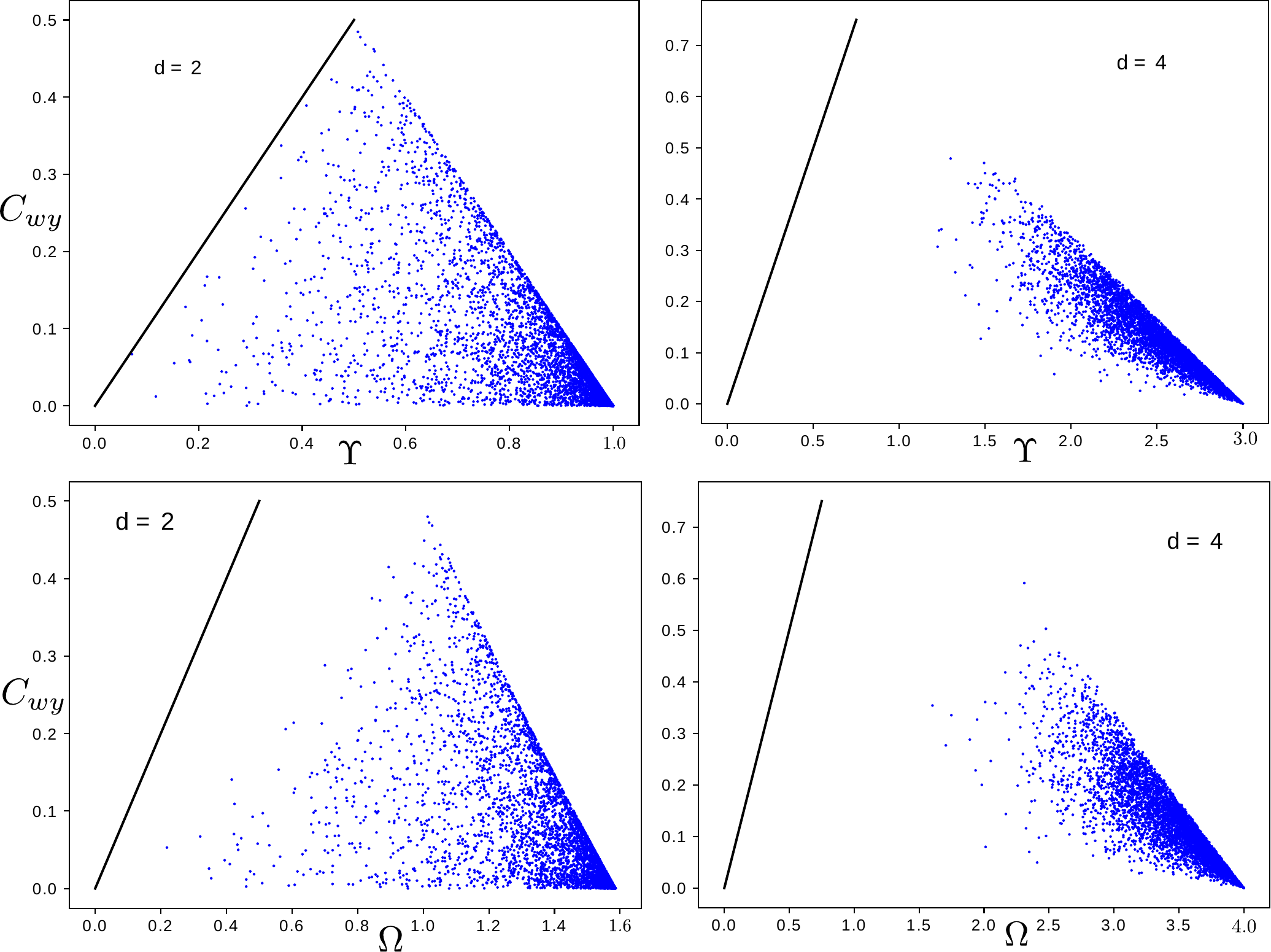}
\caption{(color online) ``Verification'' of the upper bounds for Wigner-Yanase quantum coherence in Eqs. (\ref{eq:heub}) and (\ref{eq:heub2})
for five thousand random density matrices generated for each value
of the system dimension $d$. The method used to create the random
states is the same as the one mentioned in Fig. \ref{fig:hs}. The
black lines stand for $C_{wy}(\rho)=\Upsilon$ and for $C_{wy}(\rho)=\Omega$.}
\label{fig:he}
\end{figure}

A quantum state, with spectral decomposition $\rho=\sum_{m=1}^{d}r_{m}|r_{m}\rangle\langle r_{m}|$,
is a positive matrix \cite{nielsen}, i.e., $\{r_{m}\ge0\}_{m=1}^{d}$.
So, $\sqrt{\rho}=\sum_{m=1}^{d}\sqrt{r_{m}}|r_{m}\rangle\langle r_{m}|$
is also a positive matrix. If we apply the positivity condition of
Eq. (\ref{eq:cdm2}) to $\sqrt{\rho}$ decomposed in GMB as 
\begin{equation}
    \sqrt{\rho} = \frac{\mathrm{Tr}(\sqrt{\rho})}{d}\Gamma_{0}^{d}+\frac{1}{2}\sum_{j}\langle\Gamma_{j}^{d}|\sqrt{\rho}\rangle\Gamma_{j}+\frac{1}{2}\sum_{k,l,\tau}\langle\Gamma_{k,l}^{\tau}|\sqrt{\rho}\rangle\Gamma_{k,l}^{\tau},\label{eq:sqrt}
\end{equation}
the following inequality is obtained:
\begin{eqnarray}
0 & \le & (\mathrm{Tr}(\sqrt{\rho}))^{2}-\mathrm{Tr}((\sqrt{\rho})^{2})\\
 & = & (\mathrm{Tr}(\sqrt{\rho}))^{2}\left(1-\frac{1}{d}\right)-\frac{1}{2}\sum_{j}\langle\Gamma_{j}^{d}|\sqrt{\rho}\rangle^{2}-\frac{1}{2}\sum_{k,l,\tau}\langle\Gamma_{k,l}^{\tau}|\sqrt{\rho}\rangle^{2}.
\end{eqnarray}
Using WYC in Eq. (\ref{eq:wyc}) and the linear
and von Neumann's entropies of 
\begin{equation}
    \sqrt{\rho}_{diag} := \frac{\mathrm{Tr}(\sqrt{\rho})}{d}\Gamma_{0}^{d}+\frac{1}{2}\sum_{j}\langle\Gamma_{j}^{d}|\sqrt{\rho}\rangle\Gamma_{j},
\end{equation}
we obtain the following upper bounds for WYC:
\begin{eqnarray}
C_{wy}(\rho) & \le & \left(S_{l}(\sqrt{\rho}_{diag})+(\mathrm{Tr}(\sqrt{\rho}_{diag}))^{2}-1=:\Upsilon\right)\label{eq:heub} \\
 & \le & \left(S_{vn}\left(\sqrt{\rho}_{diag}\right)+\mathrm{Tr}(\sqrt{\rho}_{diag})\left(\mathrm{Tr}(\sqrt{\rho}_{diag})-1\right)=:\Omega\right).\label{eq:heub2}
\end{eqnarray}
In the particular case of pure states, $\rho=|\psi\rangle\langle\psi|\therefore \sqrt{\rho}=\rho$,
we have $C_{wy}(\rho) = C_{hs}(\rho)$. Hence, in this case, the inequalities above are equivalent to the ones obtained for Hilbert-Schmidt's coherence in Eq. (\ref{eq:tohs}), i.e.,
\begin{equation}
C_{wy}(|\psi\rangle\langle\psi|) \le S_{l}\left(\iota^{hs}_{\rho}\right) \le S_{vn}\left(\iota^{hs}_{\rho}\right).
\end{equation}
The upper bounds for WYC
were also ``verified'' using random states, as shown in Fig. \ref{fig:he}.
Here the restrictiveness of those upper bounds is also seen to diminish
with the increase of the system dimension.

\section{Coherence-populations trade-off relations}
\label{sc:hscp}

In this section, we start rewriting the upper bound for Hilbert-Schmidt's
coherence given in Eq. (\ref{eq:hsto1}) by expressing the components
of the so called Bloch's vector corresponding to the diagonal elements
of Gell-Mann's basis, $\langle\Gamma_{j}^{d}|\rho\rangle$ with $j=1,\cdots,d-1$,
in terms of the density matrix populations, $\rho_{m,m}=\langle\beta_{m}|\rho|\beta_{m}\rangle$
with $m=1,\cdots,d$. For that purpose, after some algebraic manipulations,
one can infer that for $m=2,\cdots,d-1$:
\begin{equation}
\rho_{m,m}=\frac{1}{d}-\sqrt{\frac{m-1}{2m}}\langle\Gamma_{m-1}^{d}|\rho\rangle+\sum_{j=1}^{d-1}\frac{\langle\Gamma_{j}^{d}|\rho\rangle}{\sqrt{2j(j+1)}}\sum_{n=m}^{d-1}\delta_{n,j}.\label{eq:pop}
\end{equation}
For $m=d$ and $m=1$ we can use this same expression for the populations,
but without the last and second terms, respectively. By inverting
the expressions in Eq. (\ref{eq:pop}) iteratively, we obtain the
general expression we need to rewrite the trade-offs in Eq. (\ref{eq:tohs})
in terms of $\rho$'s populations:
\begin{equation}
\langle\Gamma_{d-j}^{d}|\rho\rangle=\sqrt{\frac{2}{(d-j+1)(d-j)}}\left(1-\sum_{n=1}^{j}(d-n+1)^{\delta_{j,n}}\rho_{d-n+1,d-n+1}\right).\label{eq:bloch}
\end{equation}
As examples, let us start considering qubit and qutrit systems. For
$d=2$, $\langle\Gamma_{1}^{d}|\rho\rangle=1-2\rho_{2,2}$ and, from
Eq. (\ref{eq:hsto1}), we get $C_{hs}(\rho)=2|\rho_{1,2}|^{2}\le2\rho_{1,1}\rho_{2,2}$,
which is equivalent to Eq. (\ref{eq:tos}). For $d=3$, $\langle\Gamma_{1}^{d}|\rho\rangle=1-\rho_{3,3}-2\rho_{2,2}$,
$\langle\Gamma_{2}^{d}|\rho\rangle=(1-3\rho_{3,3})/\sqrt{3}$ and
\begin{equation}
C_{hs}(\rho)\le2(\rho_{1,1}\rho_{2,2}+\rho_{1,1}\rho_{3,3}+\rho_{2,2}\rho_{3,3}).
\end{equation}
As this same pattern appears also for $d=4$ and for $d=5$, one would conjecture
that for any one-qudit state the following inequality will be satisfied:
\begin{equation}
C_{hs}(\rho)\le2\sum_{m=1}^{d-1}\sum_{n=m+1}^{d}\rho_{m,m}\rho_{n,n}.\label{eq:tocp}
\end{equation}

We could not give limitations for the Wigner-Yanase's coherence of a state
$\rho$ directly in terms of the density matrix populations. Notwithstanding,
relations identical to the ones above shall follow for this quantum
coherence measure if we replace $\rho$ by $\sqrt{\rho}$ in Eqs.
(\ref{eq:pop}) and (\ref{eq:bloch}) and on the right hand side of
Eq. (\ref{eq:tocp}), i.e.,
\begin{equation}
C_{wy}(\rho)\le2\sum_{m=1}^{d-1}\sum_{n=m+1}^{d}(\sqrt{\rho})_{m,m}(\sqrt{\rho})_{n,n}.\label{eq:tocpH}
\end{equation}

A simpler and general proof of this kind of inequality can be given as follows. 
It is known from Matrix Analysis \cite{horn} that all principal sub-matrices of a positive semi-definite matrix are also positive semi-definite. In particular, $A\ge\mathbb{O}\Rightarrow\begin{bmatrix} A_{j,j} & A_{j,k} \\ A_{k,j} & A_{k,k} \end{bmatrix}\ge\mathbb{O}\hspace{0.1cm}\forall j,k$. It follows then that
\begin{equation}
\sum_{j\ne k}|A_{j,k}|^{2} = \sum_{j,k}|A_{j,k}|^{2} - \sum_{j}A_{j,j}^{2}  \le \sum_{j,k}A_{j,j}A_{k,k} - \sum_{j}A_{j,j}^{2} = \sum_{j\ne k}A_{j,j}A_{k,k},
\end{equation}
from which we can obtain the inequalities in Eqs. (\ref{eq:tocp}) and (\ref{eq:tocpH}). 

In the next section, we will use one of these two expressions for giving closed formulas for measures of path information and interference pattern visibility.

\section{Quantitative complementarity relations for $d$-slits interferometers}
\label{sc:comp}

\subsection{Complementarity relations with Hilbert-Schmidt's coherence}
\label{sec:hsc}
Here we use HSC written in terms of the density matrix elements:
\begin{align} 
C_{hs}(\rho) & := \min_{\iota}||\rho-\iota||_{hs}^{2} = \min_{\iota}\sum_{j,k=1}^{d}|(\rho-\iota)_{j,k}|^{2} = \min_{\iota}\sum_{j,k=1}^{d}|\rho_{j,k}-\iota_{j}\delta_{j,k}|^{2} \\
& = \sum_{j\ne k}|\rho_{j,k}|^{2}.
\end{align}
In the sequence we show that $W:=C_{hs}$ satisfies the properties listed in the introduction. One can easily see from the expression for $C_{hs}$ that it has the properties $W1$ and $W2$, i.e., $C_{hs}$ is continuous and invariant under paths' indexes exchanges. Besides:
\begin{itemize}
\item[W3] If $\rho_{j,j}=1$ for any $j$, then all the other populations and all the coherences are null (by the restrictions $|\rho_{j,k}|^{2}\le \rho_{j,j}\rho_{k,k}\forall j,k$). Therefore $C_{hs}=0$, which is the required minimum, since we have $C_{hs}\ge 0$.
\item[W4] If $\rho_{j,j}=1/d\hspace{0.1cm}\forall j$, we shall have from $\begin{bmatrix}\rho_{j,j}&\rho_{j,k} \\ \rho_{k,j}&\rho_{k,k}\end{bmatrix}\ge\mathbb{O}$ that $|\rho_{j,k}|\le 1/d\hspace{0.1cm}\forall j\ne k$. So, for the maximum value for the density matrix coherences $\rho_{j,k}=1/d\hspace{0.1cm}\forall j\ne k$, we see that $Tr(\rho^{2})=1$ and that $C_{hs}$ reaches its maximum value, $(d-1)/d$.
\item[W5] If we set $\rho_{j,k}\rightarrow\tilde{\rho}_{j,k}=\rho_{j,k}-\epsilon$, with $\Re(\rho_{j,k})\Re(\epsilon)\ge 0$ and $\Im(\rho_{j,k})\Im(\epsilon)\ge 0$, then $|\tilde{\rho}_{j,k}|^{2}\approx|\rho_{j,k}|^{2}-2(\Re(\rho_{j,k})\Re(\epsilon)+\Im(\rho_{j,k})\Im(\epsilon))\le |\rho_{j,k}|^{2}$, which implies that $\tilde{C}_{hs}\le C_{hs}$.
\item[W6] For $0\le\omega\le 1$ and $\xi$ and $\eta$ valid density operators, we verify that $C_{hs}$ is convex as follows:
\begin{align}
& C_{hs}(\omega\xi+(1-\omega)\eta)-\omega C_{hs}(\xi)-(1-\omega)C_{hs}(\eta) \nonumber \\
&= \sum_{j\ne k}|(\omega\xi+(1-\omega)\eta)_{j,k}|^{2}-\omega\sum_{j\ne k}|\xi_{j,k}|^{2}-(1-\omega)\sum_{j\ne k}|\eta_{j,k}|^{2} \\
& = \sum_{j\ne k}\omega(\omega-1)\left((\Re(\xi_{j,k})-\Re(\eta_{j,k}))^{2}+(\Im(\xi_{j,k})-\Im(\eta_{j,k}))^{2}\right) \\ 
& \le 0.
\end{align}
\end{itemize}

Let us define
\begin{equation}
S^{\max}_{\tau}:=\max_{\rho}S_{\tau}(\rho) \text{, with } \tau=l,vn.
\end{equation}
For $d$-dimensional density matrices, $\rho=\mathbb{I}_{d}/d$ gives the maximum for the entropies:
\begin{equation}
S^{\max}_{l}=(d-1)/d \text{ and } S^{\max}_{vn}=\ln d.
\end{equation}
Now, we can rewrite the inequalities in Eqs. (\ref{eq:tohs}) as 
\begin{equation}
C_{hs}(\rho)+S^{\max}_{\tau}-S_{\tau}(\iota_{\rho}^{\tau})\le S^{\max}_{\tau}.
\end{equation}
By defining the Hilbert-Schmidt's predictability measures
\begin{align}
P^{l}_{hs}(\rho) &:=S^{\max}_{l}-S_{l}(\iota_{\rho}^{hs}) =\frac{d-1}{d}-2\sum_{m=1}^{d-1}\sum_{n=m+1}^{d}\rho_{m,m}\rho_{n,n} ,\label{Pl}\\
P^{vn}_{hs}(\rho) &:= S^{\max}_{vn}-S_{vn}(\iota_{\rho}^{hs}) = \ln d + \sum_{j=1}^{d}\rho_{n,n}\ln\rho_{n,n}, \label{Pvn}
\end{align}
we obtain the coherence-predictability complementarity relations:
\begin{equation}
C_{hs}(\rho)+P_{hs}^{\tau}(\rho)\le S^{\max}_{\tau}.
\end{equation}
One could put the upper bounds for these inequalities in the usual form, equal to one, by normalizing $C_{hs}$ and $P_{hs}^{\tau}$.

In the sequence we shall verify that the predictability measures introduced here satisfy the criteria listed in Sec. \ref{intro}:
\begin{itemize}
\item[P1] As $S_{\tau}(\iota_{\rho}^{hs})$ is a continuous function of $\{\rho_{j,j}\}_{j=1}^{d}$, so is $P^{\tau}_{hs}(\rho)$.
\item[P2] It is straightforward to see in the right hand side of Eqs. (\ref{Pl}) and (\ref{Pvn}) that $P^{l}_{hs}$ and $P^{l}_{hs}$ are invariant under exchange of paths' labels.
\item[P3] If $\rho_{j,j}=1$ for some $j$ then $\rho_{k,k}=0\hspace{0.1cm}\forall k\ne j$ and $S_{l}=1-\rho_{j,j}^{2}-\sum_{k\ne j}\rho_{k,k}^{2}=0$ and $S_{vn}=-\rho_{j,j}\ln\rho_{j,j}-\sum_{k\ne j}\rho_{k,k}\ln\rho_{k,k}=0$. So $P^{\tau}_{hs} := S^{\max}_{\tau}$.
\item[P4] If $\{\rho_{j,j}=1/d\}_{j=1}^{d}$ then $S_{l}=1-\sum_{j=1}^{d}(1/d^{2})=S^{\max}_{l}$ 
and $S_{vn}=-\sum_{j=1}^{d}(1/d)\ln(1/d)=S^{\max}_{vn}$. So $P^{\tau}_{hs}=0$. 
\item[P5] Once $P^{\tau}_{hs}$ is invariant under exchange $\rho_{j,j}\leftrightarrow\rho_{k,k}\forall j,k$ we can, without loss of generality, consider $\rho_{1,1}>\rho_{2,2}$, $\rho_{1,1}\rightarrow\rho_{1,1}-\epsilon$, 
and $\rho_{2,2}\rightarrow\rho_{2,2}+\epsilon$ for $\epsilon>0$ and $\epsilon\ll 1$. Thus
\begin{align}
\tilde{P}_{hs}^{l} &= S_{l}^{\max} - 1 + (\rho_{1,1}-\epsilon)^{2} + (\rho_{2,2}+\epsilon)^{2}+\sum_{j=3}^{d}\rho_{j,j}^{2} \\
&= S_{l}^{\max} - (1-\sum_{j=1}^{d}\rho_{j,j}^{2}) -2\epsilon(\rho_{1,1}-\rho_{2,2}) +\mathcal{O}(\epsilon^{2}) \\
&\le P_{hs}^{l}
\end{align}
and
\begin{align}
\tilde{P}_{hs}^{vn} &= S_{vn}^{\max} + (\rho_{1,1}-\epsilon)\ln(\rho_{1,1}-\epsilon) + (\rho_{2,2}+\epsilon)\ln(\rho_{2,2}+\epsilon) + \sum_{j=3}^{d}\rho_{j,j}\ln\rho_{j,j} \\
&= S_{vn}^{\max} + \sum_{j=1}^{d}\rho_{j,j}\ln\rho_{j,j} - \epsilon(\ln\rho_{1,1}-\ln\rho_{2,2}) +  (\rho_{1,1}-\epsilon)(-\epsilon/\rho_{1,1})  + (\rho_{2,2}+\epsilon)(\epsilon/\rho_{2,2}) +\mathcal{O}(\epsilon^{2})\\
&= P_{hs}^{vn} - \epsilon(\ln\rho_{1,1}-\ln\rho_{2,2}) +  \mathcal{O}(\epsilon^{2})  \\
&\le P_{hs}^{vn}.
\end{align}
Above we used $\ln(1\pm x)\approx \pm x$ for $x>0$ and $x\ll 1$.
\item[P6] The convexity of the  predictability measure $P_{hs}^{l}$ is verified as follows:
\begin{align}
& P_{hs}^{l}(\omega\xi+(1-\omega)\eta)-\omega P_{hs}^{l}(\xi)-(1-\omega)P_{hs}^{l}(\eta) \\
& = \frac{d-1}{d}-\sum_{j\ne k}(\omega\xi_{j,j}+(1-\omega)\eta_{j,j})(\omega\xi_{k,k}+(1-\omega)\eta_{k,k}) \\
& \hspace{0.4cm} - \omega\left(\frac{d-1}{d}-\sum_{j\ne k}\xi_{j,j}\xi_{k,k}\right)  -(1-\omega)\left(\frac{d-1}{d}-\sum_{j\ne k}\eta_{j,j}\eta_{k,k}\right) \\
& = \omega(1-\omega)\sum_{j\ne k}(\xi_{j,j}-\eta_{j,j})(\xi_{k,k}-\eta_{k,k}) = \omega(\omega-1)\sum_{j\ne k}(\xi_{j,j}-\eta_{j,j})^{2} \\
& \le 0.
\end{align}
For Hilbert-Schmidt's predictability quantifier $P_{hs}^{vn}$, convexity follows from the concavity of von Neumann's entropy \cite{wehrl}.
\end{itemize}

With this we have shown that although Hilbert-Schmidt distance does not provide a coherence monotone, it can be used, in conjunction with the positivity of the density matrix, to provide bona fide measures for the particle and wave aspects of a quanton in $d$-slits interferometry.

\begin{figure}
\includegraphics[scale=0.62]{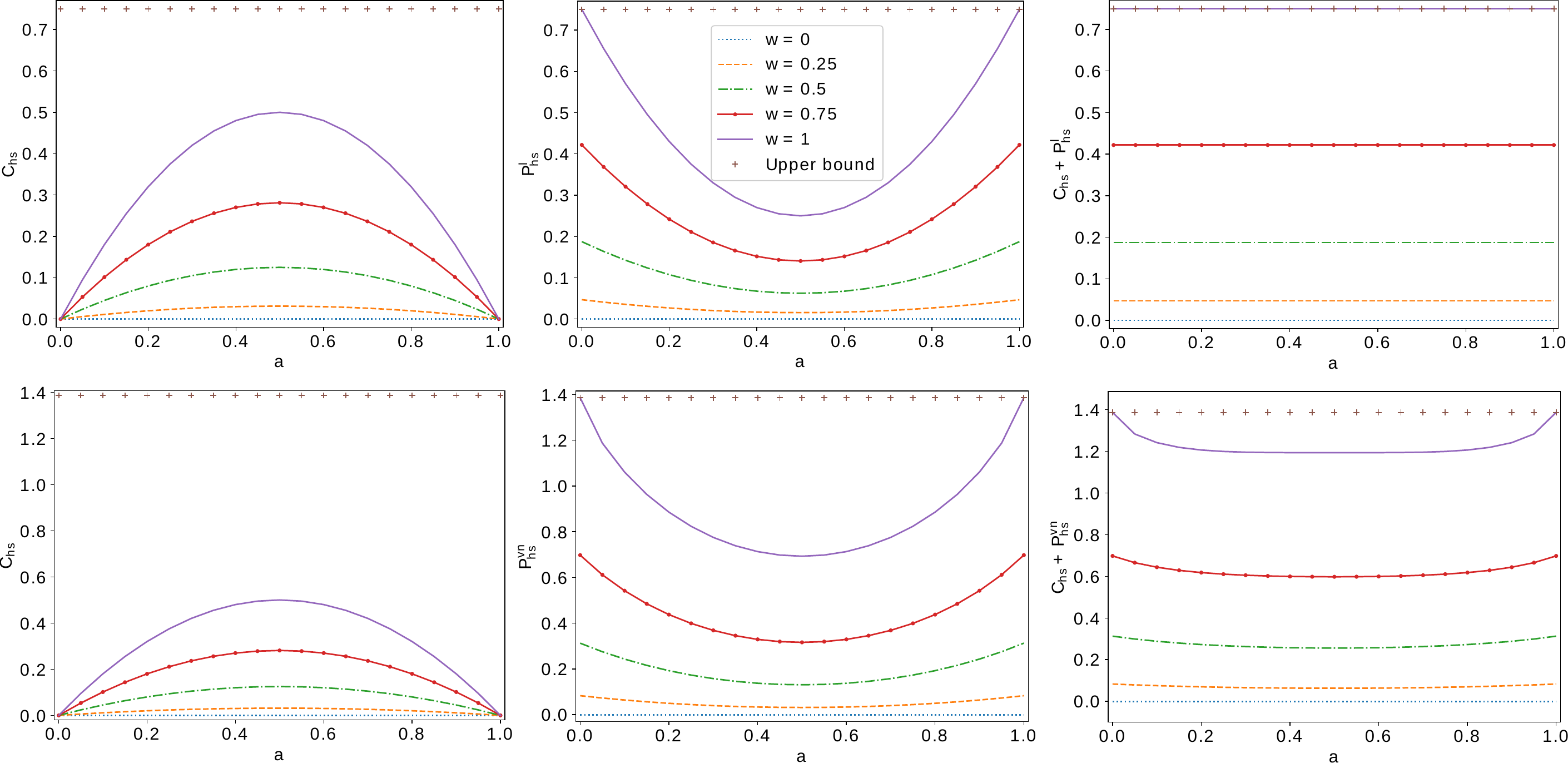}
\caption{(color online) Hilbert-Schmidt coherence function $C_{hs}$ (upper-left and lower-left plots), linear-Hilbert-Schmidt predictability $P_{hs}^{l}$ (upper-center plot), von Neumann-Hilbert-Schmidt predictability $P_{hs}^{vn}$ (lower-center plot), and the sums $C_{hs}+P_{hs}^{l}$ (upper-right plot) and $C_{hs}+P_{hs}^{vn}$ (lower-right plot) of the modified Werner's state of Eq. (\ref{eq:werner}) as a function of $a$ for some values of $w$.}
\label{fig:chs}
\end{figure}

In order to exemplify the application of our complementarity relations, we shall regard a generalized Werner's state of a ququart ($d=4$) \cite{hiroshima}:
\begin{equation}
    \rho_{w,a} = (1-w)\frac{\mathbb{I}_{4}}{4}+w|\psi\rangle\langle\psi|,
    \label{eq:werner}
\end{equation}
with $|\psi\rangle=\sqrt{a}|\beta_{0}\rangle+\sqrt{1-a}|\beta_{1}\rangle$. The Hilbert-Schmidt coherence function, the linear- and von Neumann-Hilbert-Schmidt predictability measures, their sum, and the associated upper bounds are shown graphically in Fig. \ref{fig:chs} as a function of $a$ for some values of $w$.

This figure shows that $P_{hs}^{l}$ and $P_{hs}^{vn}$ reach the respective upper bounds for $\rho_{w=1,a=1}=|\beta_{0}\rangle\langle\beta_{0}|$ and for $\rho_{w=1,a=0}=|\beta_{1}\rangle\langle\beta_{1}|$. Besides, the equality in the complementarity relation $C_{hs}+P_{hs}^{l}\le (d-1)/d$ is obtained for all $\rho_{w=1,a}$ while the inequality $C_{hs}+P_{hs}^{vn}\le \ln d$ is saturated only for $\rho_{w=1,a=1}$ and $\rho_{w=1,a=0}$.

\subsection{Complementarity relations with Wigner-Yanase's coherence}
\label{sec:wyc}

In this sub-section we will start by using the defining properties of a density matrix, $\rho\ge\mathbb{O}$ and $\mathrm{Tr}(\rho)=1$, to show that $\langle\beta_{j}|\sqrt{\rho}|\beta_{j}\rangle\ge \langle\beta_{j}|\rho|\beta_{j}\rangle$. If we define $|r_{j}\rangle:=\sum_{k=1}^{d}U_{j,k}|\beta_{k}\rangle$, from the spectral decomposition $\rho=\sum_{j=1}^{d}r_{j}|r_{j}\rangle\langle r_{j}|=\sum_{k,l=1}^{d}\left(\sum_{j=1}^{d}r_{j}U_{j,k}U_{j,l}^{*}\right)|\beta_{k}\rangle\langle\beta_{l}|$, we have $\sqrt{\rho}=\sum_{j=1}^{d}\sqrt{r_{j}}|r_{j}\rangle\langle r_{j}|=\sum_{k,l=1}^{d}\left(\sum_{j=1}^{d}\sqrt{r_{j}}U_{j,k}U_{j,l}^{*}\right)|\beta_{k}\rangle\langle\beta_{l}|$ and thus
\begin{equation}
(\sqrt{\rho})_{j,j}=\sum_{k}\sqrt{r_{k}}|U_{k,j}|^{2}\ge\sum_{k}r_{k}|U_{k,j}|^{2}=\rho_{j,j}.
\end{equation}
So
\begin{align}
C_{wy}(\rho) & = 1-\sum_{j=1}^{d}\langle\beta_{j}|\sqrt{\rho}|\beta_{j}\rangle^{2} \\ 
& \le 1-\sum_{j=1}^{d}\langle\beta_{j}|\rho|\beta_{j}\rangle^{2} = S_{l}(\rho_{diag}) \le S_{vn}(\rho_{diag}),
\end{align}
with $\rho_{diag}=diag(\rho_{1,1},\rho_{2,2},\cdots,\rho_{d,d})$ and the linear and von-Neumann  entropies are defined in Sec. \ref{sec:tohs}. Thus we identify the complementarity relations:
\begin{equation}
C_{wy}(\rho)+P_{hs}^{\tau}(\rho)\le S_{\tau}^{\max},
\label{eq:cpwy}
\end{equation}
with $\tau=l,vn$ and with the predictability measures complementary to Wigner-Yanase's coherence being the same as for Hilbert-Schmidt's coherence, since $\rho_{diag} = \iota^{hs}_{\rho}$. These functions appear, together with $S_{\tau}^{\max}$, in Sec. \ref{sec:hsc}. We observe that other candidate predictability measures $P_{wy}$ could be defined using the inequalities in Eqs. (\ref{eq:heub}) and (\ref{eq:heub2}). But we do not include such functions here because for them we succeed in verifying axioms P1-P6 only in terms of changes in $\sqrt{\rho}_{diag}$, which still lacks physical significance. 

In the sequence we verify that Wigner-Yanase's coherence satisfies the properties listed in the Introduction for a measure of the wave character of a quanton:
\begin{itemize}
\item[W1] Continuity if $C_{wy}$ follows from the continuity of $\{(\sqrt{\rho})_{j,j}\}_{j=1}^{d}$.
\item[W2] Once $\sum_{j=1}^{d}((\sqrt{\rho})_{j,j})^{2}$ does not change under $|\beta_{j}\rangle\leftrightarrow|\beta_{k}\rangle$, $C_{wy}$ is invariant under paths' indexes exchanges.
\item[W3] If $\rho_{j,j}=1$ for some $j$, then by $\mathrm{Tr}(\rho)=1$ we have to have $\rho_{k,k}=0\hspace{0.1cm}\forall k\ne j$, which implies that $\rho=|\beta_{1}\rangle\langle\beta_{1}|=\sqrt{\rho}$. Therefore $(\sqrt{\rho})_{j,j}=1 \text{ and }(\sqrt{\rho})_{k,k}=0\hspace{0.1cm}\forall k\ne j$. So $C_{wy}=1-1=0$, which is its minimum value.
\item[W4] If $\rho$ is a pure state then $\sqrt{\rho}=\rho$. Therefore, if $\{\rho_{j,j}=1/d\})_{j=1}^{d}$ then $C_{wy}=(d-1)/d$, which is its maximum value.
\item[W5] We can diminish $|\rho_{j,k}|$ infinitesimally by taking $\rho_{j,k}\rightarrow (1-\epsilon)\rho_{j,k}$, with $\epsilon\in\mathbb{R}_{+}$ and $\epsilon\ll 1$. By noticing that $\rho_{j,k}=\sum_{l}(\sqrt{\rho})_{j,l}(\sqrt{\rho})_{l,k}$, we see that this change leads to $\rho_{j,k}\rightarrow\sum_{l}((1-\epsilon)(\sqrt{\rho})_{j,l})(\sqrt{\rho})_{l,k}$, which is equivalent to multiplying the $j$-th row of $\sqrt{\rho}$ by $1-\epsilon$. Thus it follows that
\begin{align}
    \tilde{C}_{wy} & = \sum_{l\ne j}|(1-\epsilon)(\sqrt{\rho})_{j,l}|^{2} + \sum_{k\ne j}\sum_{l\ne k}|(\sqrt{\rho})_{k,l}|^{2} \\
    & \approx(1-2\epsilon)\sum_{l\ne j}|(\sqrt{\rho})_{j,l}|^{2} + \sum_{k\ne j}\sum_{l\ne k}|(\sqrt{\rho})_{k,l}|^{2} \\
    & = C_{wy}(\rho) - 2\epsilon\sum_{l\ne j}|(\sqrt{\rho})_{j,l}|^{2} \\
    & \le C_{wy}(\rho).
\end{align}
\item[W6] Convexity of $C_{wy}$ follows from the convexity of Wigner-Yanase's skew information $I_{wy}$ \cite{wigner}.
\end{itemize}

\begin{figure}
\includegraphics[scale=0.57]{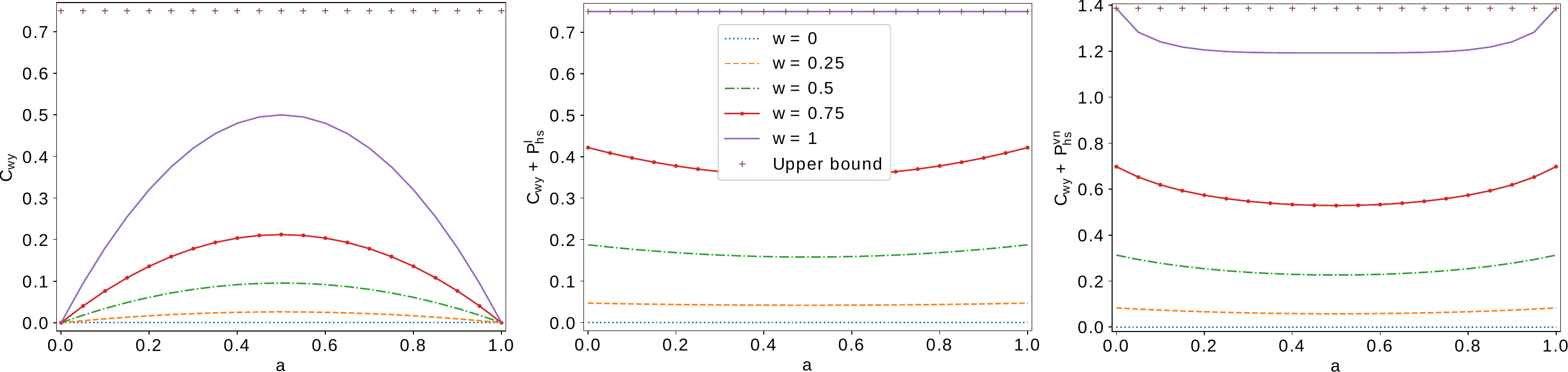}
\caption{(color online) Wigner-Yanase's coherence (left plot), $C_{wy}+P_{hs}^{l}$ (center plot), and $C_{wy}+P_{hs}^{vn}$ (right plot) for the modified Werner's state of Eq. (\ref{eq:werner}) as a function of $a$ for some values of $w$.}
\label{fig:cwy}
\end{figure}

In Fig. \ref{fig:cwy} we exemplify the application of the complementarity relations of Eq. (\ref{eq:cpwy}) for the state of Eq. (\ref{eq:werner}). The general aspects of the obtained results are similar to those described above for the Hilbert-Schmidt coherence.

\subsection{Complementarity relations with $l_{1}$-norm coherence}
\label{sec:l1c}
In this subsection we derive quantitative complementarity relations for $d$- dimensional systems by applying $l_{1}$-norm coherence \cite{baumgratz}:
\begin{align}
    C_{l_{1}}(\rho) &= \min_{\iota}||\rho-\iota||_{l_{1}} = \min_{\iota}\sum_{j,k=1}^{d}|(\rho-\iota)_{j,k}| = \min_{\iota}\sum_{j,k=1}^{d}|\rho_{j,k}-\iota_{j}\delta_{j,k}| \\
    & =\sum_{j\ne k}|\rho_{j,k}|.
\end{align}
Here we will use again $\rho\ge\mathbb{O}\Rightarrow |\rho_{j,k}|^{2}\le\rho_{j,j}\rho_{k,k}\hspace{0.1cm}\forall j,k$ to obtain
\begin{equation}
    C_{l_{1}}(\rho) \le \sum_{j\ne k}\sqrt{\rho_{j,j}\rho_{k,k}}\le d-1,
\end{equation}
from which follows the complementarity relation:
\begin{equation}
    C_{l_{1}}(\rho)+P_{l_{1}}(\rho) \le d-1,
    \label{eq:cpl1}
\end{equation}
with the $l_{1}$-norm predictability defined as:
\begin{align}
    P_{l_{1}}(\rho) & := d-1 - \sum_{j\ne k}\sqrt{\rho_{j,j}\rho_{k,k}} \\
    & = d-1 - 2\sum_{j < k}\sqrt{\rho_{j,j}\rho_{k,k}}.
\end{align}

It worthwhile mentioning that for $d = 2$ we can write $P^l_{hs}(\rho) = (\rho_{1,1} - \rho_{2,2})^2$, which is similar to predictability measure used in \cite{dur, englert}. But we notice that $P = (f(\rho_{1,1}) - f(\rho_{2,2}))^2$ is also a bona-fide measure of predictability, with $f$ being any monotonic increasing function of the probabilities $\rho_{j,j}, \  j= 1, 2$. Hence, for $f(x) =  \sqrt{x}$ the $l_{1}$-norm predictability is a generalization of two dimensional function $P = (\sqrt{\rho_{1,1}} - \sqrt{\rho_{2,2}})^2$.

Next we verify that $C_{l_{1}}$ satisfy the axioms for a measure of the wave aspect of a quanton:
\begin{itemize}
    \item[W1] Continuity follows from the continuity of the absolute value function.
    \item[W2] Invariance under paths' indexes exchange follows directly from the analytical expression for $C_{l_{1}}$.
    \item[W3] If $\rho_{j,j}=1$ for some $j$, then $\rho_{k,k}=0\hspace{0.1cm}\forall k\ne j$ and, by $|\rho_{j,k}|^{2}\le\rho_{j,j}\rho_{k,k}$, $\rho_{j,k}=0\hspace{0.1cm}\forall j\ne k$. Therefore $C_{l_{1}}=0$.
    \item[W4] If $\{\rho_{j,j} = 1/d \}_{j=1}^{d}$, then the same inequality used to prove W3 leads to $|\rho_{j,k}|\le 1/d$. The equality gives $\mathrm{Tr}(\rho^{2})=1$ and $C_{l_{1}}=d-1$, which is its maximum value.
    \item[W5] For $\epsilon\in\mathbb{C}$, $|\epsilon|\ll 1$, $\Re(\rho_{j,k})\Re(\epsilon)>0$, $\Im(\rho_{j,k})\Im(\epsilon)>0$, we set $\tilde{\rho}_{j,k}=\rho_{j,k}-\epsilon$. Then
    \begin{align}
        |\tilde{\rho}_{j,k}| & = \sqrt{(\rho_{j,k}-\epsilon)(\rho_{j,k}^{*}-\epsilon^{*})} \\ 
        &\approx \sqrt{|\rho_{j,k}|^{2}-2\Re(\rho_{j,k}\epsilon^{*})} \\ 
        & \approx |\rho_{j,k}|\left(1-\Re(\rho_{j,k}\epsilon^{*})/|\rho_{j,k}|^{2}\right),
    \end{align}
    which gives $\tilde{C}_{l_{1}}=C_{l_{1}}-\Re(\rho_{j,k}\epsilon^{*})/|\rho_{j,k}|\le C_{l_{1}}$.
    \item[W6] For $0\le\omega\le 1$ and $\xi$ and $\eta$ valid density operators, we verify convexity of $C_{l_{1}}$ as follows:
    \begin{align}
        C_{l_{1}}(\omega\xi+(1-\omega)\eta) & = \sum_{j\ne k}|(\omega\xi+(1-\omega)\eta)_{j,k}| \\ 
        & = \sum_{j\ne k}|\omega\xi_{j,k}+(1-\omega)\eta_{j,k}| \\
        & \le \sum_{j\ne k}(|\omega\xi_{j,k}|+|(1-\omega)\eta_{j,k}|)\\
        &= \omega C_{l_{1}}(\xi)+(1-\omega)C_{l_{1}}(\eta).
    \end{align}
\end{itemize}

At last we verify that $P_{l_{1}}$ satisfies the axioms listed in Sec. \ref{intro} for a measure of predictability:
\begin{itemize}
    \item[P1] Continuity of $P_{l_{1}}$ follows from the continuity of the square root.
    \item[P2] The sum $\sum_{j\ne k}\sqrt{\rho_{j,j}\rho_{k,k}}$ warrants invariance under paths' indexes exchanges.
    \item[P3] If $\rho_{j,j}=1$ for some $j$, then $\rho_{k,k}=0\hspace{0.1cm}\forall k\ne j$. Thus $P_{l_{1}} = d-1 - 0$, which is its maximum value.
    \item[P4] If $\{\rho_{j,j}=1/d\}_{j=1}^{d}$, then $\sum_{j\ne k}\sqrt{\rho_{j,j}\rho_{k,k}}=d-1$, and $P_{l_{1}}=0$, which is its minimum value.
    \item[P5] In view of $P_{2}$, we set $\rho_{1,1}>\rho_{2,2}$, $\rho_{1,1}\rightarrow\rho_{1,1}-\epsilon$, and $\rho_{2,2}\rightarrow\rho_{2,2}+\epsilon$ with $\epsilon\in\mathbb{R}_{+}$ and $\epsilon\ll 1$. So
    \begin{align}
        \tilde{P}_{l_{1}} & = d-1-2\sqrt{\rho_{1,1}-\epsilon}\sqrt{\rho_{2,2}+\epsilon}-2\sqrt{\rho_{1,1}-\epsilon}\sum_{k=3}^{d}\sqrt{\rho_{k,k}} \nonumber \\ 
        & \hspace{0.3cm}-2\sqrt{\rho_{2,2}+\epsilon}\sum_{k=3}^{d}\sqrt{\rho_{k,k}} -2\sum_{j=3}^{d-1}\sum_{k=j+1}^{d}\sqrt{\rho_{j,j}\rho_{k,k}} \\
        & \approx d-1-2\sqrt{\rho_{1,1}\rho_{2,2}}\left(1-\epsilon/2\rho_{1,1}\right)\left(1+\epsilon/2\rho_{2,2}\right)-2\sqrt{\rho_{1,1}}\left(1-\epsilon/2\rho_{1,1}\right)\sum_{k=3}^{d}\sqrt{\rho_{k,k}} \nonumber \\
        & \hspace{0.3cm} -2\sqrt{\rho_{2,2}}\left(1+\epsilon/2\rho_{2,2}\right)\sum_{k=3}^{d}\sqrt{\rho_{k,k}} -2\sum_{j=3}^{d-1}\sum_{k=j+1}^{d}\sqrt{\rho_{j,j}\rho_{k,k}} \\
        & \approx P_{l_{1}} - \epsilon\left(\sqrt{\frac{\rho_{1,1}}{\rho_{2,2}}}-\sqrt{\frac{\rho_{2,2}}{\rho_{1,1}}}\right) - \epsilon\left(\frac{1}{\sqrt{\rho_{2,2}}}-\frac{1}{\sqrt{\rho_{1,1}}}\right)\sum_{k=3}^{d}\sqrt{\rho_{k,k}} \\
        & < P_{l_{1}}.
    \end{align}
    \item[P6] We will prove convexity through the positivity of the Hessian matrix $(H_{n,m})=(\partial_{n}\partial_{m}f)$, with $\partial_{n}:=\frac{\partial}{\partial x_{n}}$ and $f=\alpha-\sum_{j\ne k}\sqrt{x_{j}x_{k}}$ where $\alpha$ is a constant, i.e., we will verify that $\langle y|H|y\rangle=\sum_{n,m}y_{n}^{*}y_{m}H_{n,m} \ge 0 \hspace{0.2cm} \forall|y\rangle\in\mathbb{C}^{d}$. Once the diagonal and off-diagonal elements of $H$ are given, respectively, by: $\partial_{m}\partial_{m}f=\frac{1}{2}\sum_{j\ne m}\sqrt{x_{j}/x_{m}^{3}}$ and $\partial_{n}\partial_{m}f=-1/2\sqrt{x_{n}x_{m}}$, we shall have:
    \begin{align}
        \langle y|H|y\rangle & = \sum_{m}|y_{m}|^{2}\frac{1}{2}\sum_{n\ne m}\sqrt{\frac{x_{n}}{x_{m}^{3}}} + \sum_{n\neq m}y_{n}^{*}y_{m}\frac{-1}{2\sqrt{x_{n}x_{m}}} \\
        & = \frac{1}{4}\sum_{m\ne n}\left(\frac{|y_{m}|^{2}x_{n}^{1/2}}{x_{m}^{3/2}} + \frac{|y_{n}|^{2}x_{m}^{1/2}}{x_{n}^{3/2}} - \frac{y_{n}^{*}y_{m}+y_{m}^{*}y_{n}}{\sqrt{x_{n}x_{m}}}\right) \\
        & = \frac{1}{4}\sum_{m\ne n}x_{n}^{1/2}x_{m}^{1/2}\left(\frac{|y_{m}|^{2}}{x_{m}^{2}} + \frac{|y_{n}|^{2}}{x_{n}^{2}} - \frac{y_{n}^{*}y_{m}+y_{m}^{*}y_{n}}{x_{n}x_{m}}\right) \\
        & = \frac{1}{4}\sum_{m\ne n}x_{n}^{1/2}x_{m}^{1/2}\left|\frac{y_{m}}{x_{m}}-\frac{y_{n}}{x_{n}}\right|^{2} \ge 0.
    \end{align}
\end{itemize}

\begin{figure}
\includegraphics[scale=0.58]{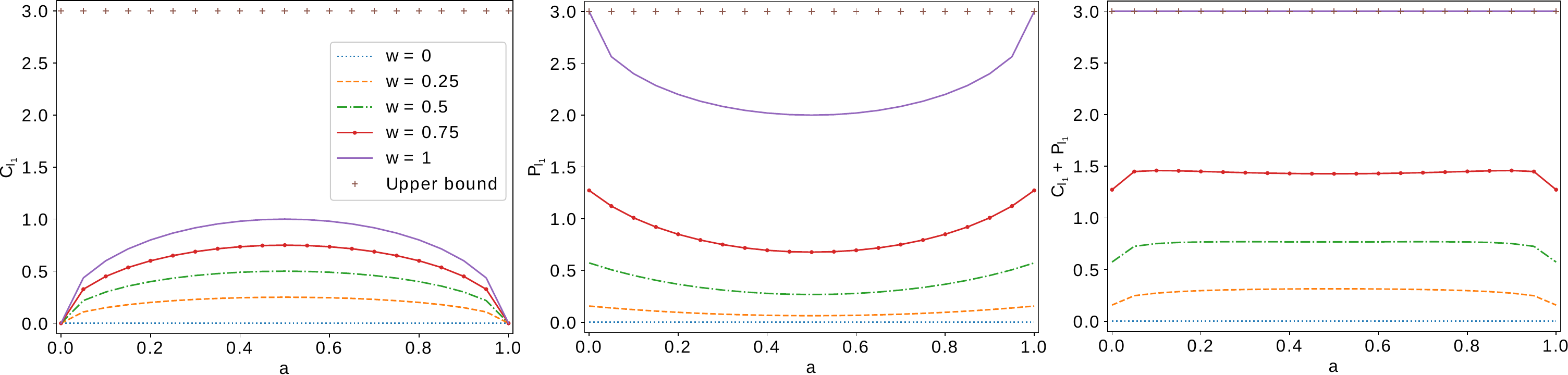}
\caption{(color online) $l_{1}$-norm coherence (left plot), $l_{1}$-norm predictability (center plot), and $C_{l_{1}}+P_{l_{1}}$ (right plot) for the modified Werner's state of Eq. (\ref{eq:werner}) as a function of $a$ for some values of $w$.}
\label{fig:cl1}
\end{figure}

In Fig. \ref{fig:cl1} we instantiate the application of the inequality of Eq. (\ref{eq:cpl1}) for the quantum state in Eq. (\ref{eq:werner}). We observe that although $C_{l_{1}}$ does not reach the upper bound and $P_{l_{1}}$ does reach this value only for $\rho_{w=1,a=0}$ and for $\rho_{w=1,a=1}$, the coherence-predictability relation is saturated for all $\rho_{w=1,a}$.

\section{Conclusions}

\label{sec:conc}

Quantum coherence (QC) is an important resource in Quantum Information
Science \cite{hillery,kammerlander,streltsov_sm,yu,yuan,shi,giorda,zhang,Pozzobom2017a,buruaga,li_epjd,brandner,scholes,bengtson,pinto,southwell,li_pla}.
In this article we proved upper bounds for Hilbert-Schmidt's QC of a general one-qudit state $\rho$ by its associated
incoherent uncertainty measured using the linear entropy and von Neumann's
entropy of the closest incoherent mixture. Similar bounds were obtained for Wigner-Yanase QC in terms of entropies of the diagonal part of $\sqrt{\rho}$. We
also wrote these inequalities with the upper bound given in terms of the populations of
the density matrix or of its square root. 

We have presented numerical examples of the proven inequalities using
random quantum states. These examples showed that the given upper
bounds are tight for qubits and that they have their restrictiveness
progressively weakened as the system dimension grows. So, in the future
it would be interesting to investigate if the positivity of coefficients
in Eq. (\ref{eq:coeff}) others than the one considered in this article
(see e.g. \cite{byrd,kimura}) may be used to obtain similar but more
generally stronger upper bounds for quantum coherence. 

We showed that our inequalities can be used to derive quantitative wave-particle duality relations. 
In our formalism these relations appear naturally, with the predictability measures 
defined by the inequalities themselves, which, by its turn, follows directly from the positivity of the density matrix (akin to what was implicitly done for 2-slits interferometers in Ref. \cite{englert_prl}). 
Finding another applications for the reported inequalities is another natural
continuation for the present research. One possibility for investigation
is regarding coherence generation via quantum operations with restrictions
on the possible density matrix populations changes \cite{brandao_rt,chitambar_rt}.
Other promising candidate area for application of quantum coherence--incoherent
uncertainty trade-off relations reported here is quantum thermodynamics
\cite{huber,goold,anders,lostaglio}. In this scenario, if the reference
basis is the energy basis, restrictions on populations changes shall
be related to restrictions on energy changes. And these restrictions
may be useful for analyzing thermodynamical processes that consume
or create quantum coherence.

\begin{acknowledgments}
This work was supported by the Coordena\c{c}\~ao de Aperfei\c{c}oamento de Pessoal de N\'ivel Superior (CAPES), process 88882.427924/2019-01, by the Funda\c{c}\~{a}o de Amparo \`{a} Pesquisa do Estado do Rio Grande do Sul (FAPERGS), and by the Instituto Nacional de Ci\^encia e Tecnologia de Informa\c{c}\~ao Qu\^antica (INCT-IQ), process 465469/2014-0.
\end{acknowledgments}

\end{document}